\documentclass[10pt,twocolumn]{article}
\usepackage{times}
\usepackage{epsfig}
\usepackage{graphicx}
\usepackage{algorithm2e}
\usepackage{amsmath}
\usepackage{abstract}
\usepackage{soul} 
\usepackage{url}
\usepackage{color}
\usepackage{ulem}

\newcommand{\nchoosek}[2]{\left(\begin{array}{c}#1\\#2\end{array}\right)}

\newcommand{\Section}[1]{\vspace{-8pt}\section{\hskip -1em.~~#1}\vspace{-3pt}}
\newcommand{\SubSection}[1]{\vspace{-3pt}\subsection{\hskip -1em.~~#1}\vspace{-3pt}}

\begin{document}

\title{\Large\bf Evolving Clustered Random Networks}

\author{Shweta Bansal \footnotemark[1] \and Shashank Khandelwal \and Lauren Ancel Meyers}

\twocolumn[

\maketitle

\begin{onecolabstract}
{We propose a Markov chain simulation method to generate simple
connected random graphs with a specified degree sequence and level
of clustering. The networks generated by our algorithm are random in
all other respects and can thus serve as generic models for studying
 the impacts of degree distributions and clustering on dynamical processes
 as well as null models for detecting other structural properties in empirical networks.}
\end{onecolabstract}
\vspace{0.35in}]

\vspace{-0.25in}
\begin{table*}[t]
\begin{center}
\begin{tabular}{|c|c|c|c|c|c|c|c|}
\hline
\\[0.25pt]
Empirical Network&$N$&$<d>$&$<d^2>$&$C$&$T$&$\tilde{C}$&$\tilde{T}$\\
\hline
Vancouver Urban Contacts&2627&13.9&265&0.07&0.09&0.09&0.14\\
WWW Subgraph&4271&4.2&119&0.09&0.01&0.15&0.08\\
Yeast Protein Interactions&4713&6.3&152&0.13&0.06&0.14&0.18\\
Astro-Phys Collaborations&5973&4.1&35&0.51&0.38&0.60&0.62\\
US Air Traffic&165&38.0&2765&0.86&0.58&0.97&0.96\\
\hline
\end{tabular}
\caption{The number of nodes ($N$), the average node degree ($<d>$),
the mean-squared of node degree ($<d^2>$), clustering coefficient
($C$), transitivity ($T$), Soffer-Vasquez clustering coefficient
($\tilde{C}$), and Soffer-Vasquez transitivity ($\tilde{T}$) for a set of
empirical networks.}
\end{center}
\end{table*}

\Section{Introduction}

\footnotetext[1]{Corresponding Author: shweta@sbansal.com}

Complex networks such as those formed by the links of the World Wide Web, social
contacts between individuals in a city, and transportation routes
have received much attention in the last decade. Recent studies have sought to characterize and explain non-trivial structural properties
such as heavy-tail degree distributions, clustering, short
average path lengths, degree correlations and community structure.
These properties appear in diverse natural and manmade systems, and can
fundamentally influence dynamical processes on these networks
\cite{watts, meyers_sars, newman_watts, albert, faloutsos,
keeling_networks, albert_stat, albert_error}.

Clustering, a property describing the presence of triangles in a network, is an important
topological characteristic that can significantly impact dynamical processes over complex networks
\cite{watts, mejn_clust, serrano, soffer, petermann, keeling_invasion}. 
It is often correlated with local graph properties such
as correlations in the number of edges emanating from neighboring
vertices \cite{serrano}, as well as global properties such as motifs
\cite{milo2, vazquez2} and community structure \cite{radicchi}.

\par
Random graphs are graphs that are generated by some random process \cite{mejn_random}.
They are widely used as models of complex networks
\cite{newman_watts} and can assume various levels of complexity. The simplest model for generating random graphs, with only a single parameter, is the Bernoulli
or Erd\"{o}s-Renyi random graph model, which produces graphs that are
completely defined by their average degree and are random in all
other respects.  A slightly more complex and general model is one that generates graphs with a specified degree distribution (or degree sequence) and are random in all other respects.  These models can be extended to include additional structural constraints, such as degree correlations or the density of triangles or longer cycles. Here, we define a
a random graph model which is constrained by the node
degree distribution and the density of triangles in the graph.

\vspace{-0.2in} \SubSection{Clustering in Real Networks}

Clustering in real networks can stem from two sources: (a) it can
arise as a byproduct of other, more fundamental, topological properties such as the
degree sequence (distribution) or degree correlations; or (b) it can
be generated directly by some inherent property or mechanism
within the system, for example, ``the friends of my friends tend to
become my friends" in social networks.

Some researchers have claimed that high clustering is a general feature of complex
networks \cite{serrano}. When we measured clustering in a variety of
empirical technological, biological and social networks, however, 
we found that it varies considerably. Table 1 shows that the
clustering coefficients and transitivity values (a local and global measure
of clustering, respectively) for these networks span the entire range of possible values (zero to one). Thus, it is important to understand not only the origins of clustering, but also the impact of clustering on network functions and dynamics. Towards this end, we introduce a method for generating random networks with a specified level of clustering.

\SubSection{Previous Work \& Motivation}

The study of clustering in complex networks began with the seminal
work of Watts and Strogatz \cite{watts}.  The authors presented a
graph model with high clustering and low average path length, now
known as the \textit{small-world property}.  Although not intended as
a generative algorithm for clustered graphs, the model produces
graphs with clustering spanning the range from 0 to 1. The graphs
generated under this model, however, have rigid spatial structure
and cannot accommodate varying degree distributions.

\par
The first algorithms to explicitly generate graphs with a specified
level of clustering for arbitrary degree distributions belonged to
the class of projected bipartite graphs.  Newman \cite{mejn_clust}
introduced a three-step method that first builds a bipartite graph of individuals and affiliations, then projects the bipartite graph to a unipartite graph of individuals only, and finally runs a percolation process over the unipartite graph.
This results in a clustered graph with a degree
distribution that depends on the original distributions of numbers of individuals per group and groups per individual. The level of clustering in the final graph varies 
smoothly from 0 to 1 as a function of the percolation probability. In
\cite{guil}, Guillaume suggested a similar bipartite graph approach.
Although these approaches can generate clustered graphs with diverse degree distributions, they lack straightforward methods for choosing parameters that yield graphs with not only a pre-specified clustering coefficient but also a pre-specified degree distribution. These algorithms also tends to produce disconnected graphs that leave a significant proportion of the graph vertices isolated.

\par
A second class of clustered graph models use ''growing network`` algorithms
\cite{boguna, volz_clust,trapman}. The inputs to these models are a
degree distribution and level of clustering. The method begins with a
set of vertices with no edges; the graph is then ``grown" by adding
edges based on the degree and clustering constraints. Although the
algorithms of this class allow for arbitrary degree distributions
and levels of clustering, they either require a complex
implementation \cite{boguna}, produce graphs of a highly specific
structure \cite{trapman} or introduce large amounts of degree correlations
\cite{trapman,volz_clust}.

\par
Here, we present a model that generates simple and connected graphs with
prescribed degree sequences and a specified frequency of triangles,
while maintaining a graph structure that is as random (uncorrelated) as
possible. There is an important difference between our model and
previous work in the area. Prior models were intended to generate
clustered graphs that replicate the properties of real-world
networks; our goal is to generate a class of null networks with arbitrary degree
distributions that are simple and connected and have a high density of triangles, 
but are random in all other respects.

\par
Such a method is useful for two primary reasons: First, network
structure fundamentally influences the functions of and 
dynamical processes on networks. We can use random clustered graphs 
to study the consequences of clustering, both independently and
in combination with various degree patterns. Second, these networks
can serve as null models for detecting whether an empirical network can be boiled down to its degree distribution and clustering values or, instead, contains substantial 
degree correlations or other important structures (beyond the byproducts of the degree distribution and clustering). One would first use the algorithm to generate an ensemble networks that match the empirical degree distribution and clustering values, and then compare the structural, functional, or dynamical properties of the empirical network to those of the random networks. 

\par
In Section 2, we review common measures of clustering and introduce our
Markov chain model and algorithm for generating clustered
graphs with a specified degree sequence. In Section 3, we test
our algorithm with numerical simulations and discuss the structural
properties of the generated graphs. Finally, in Section 4, we use
the generated graphs to detect deviation from randomness in
empirical networks.


\vspace{-0.25in}

\begin{figure*}[t]
\begin{center}
\includegraphics[width=12.5cm]{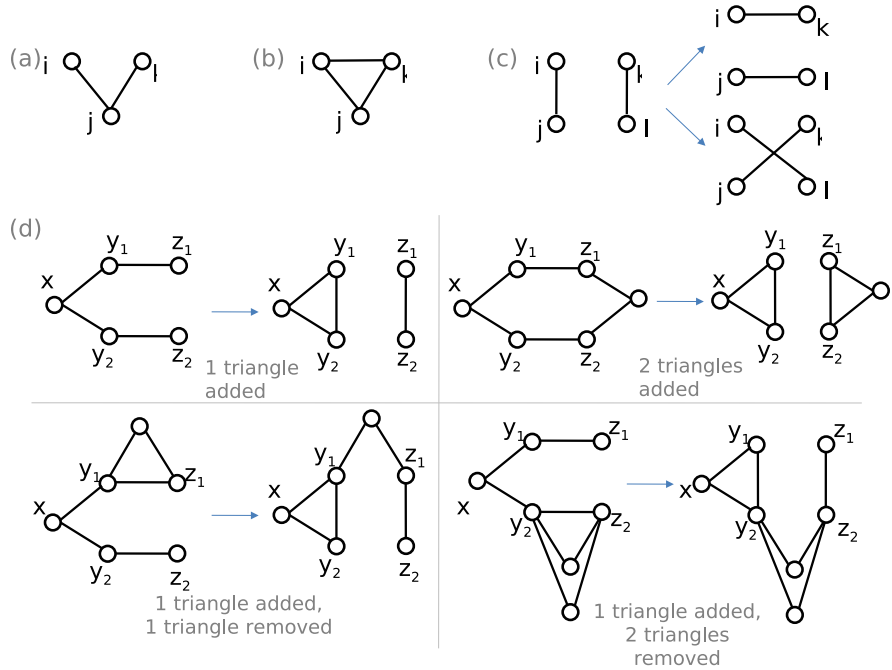}
\end{center}
\caption{(a) a triple among the nodes $i,j,k$ (b) a triangle among
the nodes $i,j,k$ (c) A rewiring of edges $(i,j)$ and $(k,l)$ can
result in $(i,k)$ and $(j,l)$, or $(i,l)$ and $j,k)$ (d) Four (among
many) scenarios for the result of one rewiring step of our
algorithm. The configuration of edges before (left) and after
(right) a rewiring step are shown for each scenario. The two bottom scenarios would be rejected by our algorithm as they do not strictly increase the number of triangles.}
\end{figure*}

\Section{Methods and Model}

Our random graph generation method begins with a random graph and iteratively rewires edges to introduce triangles. Network rewiring is a well-known
method for generating networks with desired properties \cite{milo}.
Two edges are called \textit{adjacent} if they connect to a common node. Each \textit{rewiring} is performed on
two non-adjacent edges of the graph and consists of removing these
two edges and replacing them with another pair of edges.
Specifically, a pair of edges $(i,j)$ and $(k,l)$ is replaced with
either $(i,k)$ and $(j,l)$, or $(i,l)$ and $(j,k)$ (as illustrated
in Figure 1c). This change in the graph leaves the degrees of the
participating nodes unchanged, thus maintaining the specified degree
sequence. Below we describe a rewiring 
algorithm that increases the level of clustering in a random graph, while preserving the degree sequence.

\vspace{-0.25in}
\begin{figure*}[t]
\begin{center}
\includegraphics[width=12.5cm]{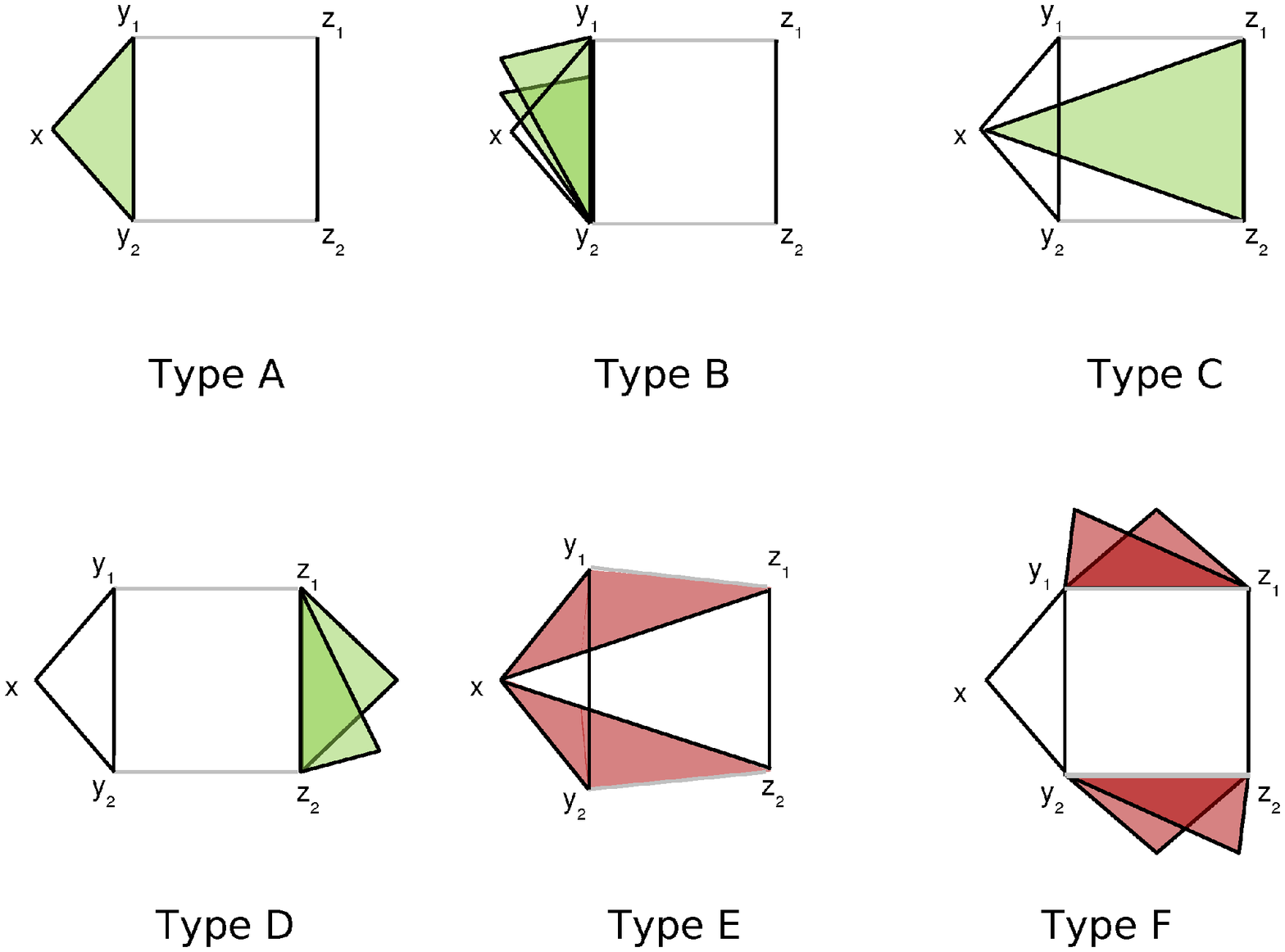} \caption{Possible triangle additions
(green) and removals (red) in one step of the rewiring procedure.
Black lines represent existing edges and edges added after a
rewiring event, gray lines represent edges lost during a rewiring
event.}
\end{center}
\end{figure*}

\SubSection{Measures of Clustering}

We begin with a graph $G=(V,E)$ which is undirected and
simple (no self-loops or multiple edges).  $V$ is the set of
vertices of $G$ and $E$ is the set of the edges.  We let $N=|V|$ and $M=|E|$ denote
the number of nodes and edges in $G$, respectively. 
The \textit{degree} of a node $i$ will be denoted $d_i$.
The set of degrees for all nodes in the graph makes up the
\textit{degree sequence}, which follows a probability distribution
called the \textit{degree distribution}.

\par

Clustering is the likelihood that two neighbors of a given node are
themselves connected.  In terms of social networks, it measures the
probability that ``the friend of my friend is also my friend.''  In
topological terms, clustering measures the density of \textit{triangles}
in the graph, where a triangle is the existence of the set of edges
${(i,j), (i,k), (j,k)}$ between any triplet of nodes $i,j,k$ (Figure 1b). 

To quantify the local presence of triangles, 
we define $\delta(i)$ as the number of
triangles in which node $i$ participates.  Since each triangle
consists of three nodes, it is counted thrice when we sum
$\delta(i)$ for each node in the graph. Thus the total number of triangles
in the graph is
$$\delta(G) = 1/3 \sum_{i \in V} \delta(i).$$

A \textit{triple} is a set of three nodes, $i,j,k$
that are connected by edges $(i,j)$ and $(j,k)$, regardless of the
existence of the edge $(i,k)$ (Figure 1a). The number of
triples of node $i$ is simply $\tau(i) = \nchoosek{d_i}{2}$, assuming
$d_i \geq 2$. To compute the total number of triples in the graph we
sum $\tau(i)$:
$$\tau(G) = \sum_{i \in V} \tau(i).$$ 
The term \textit{triadic closure} refers the conversion of a triple into a triangle via the addition of a third edge [INSERT REFS].

The \textit{clustering coefficient} was introduced by Watts and
Strogatz \cite{watts} as a local measure of triadic closure.  For a
node $i$ with $d_i \geq 2$, the clustering coefficient $c(i)$ is the
fraction of triples for node $i$ which are closed, and can be
measured as $\delta(i)/\tau(i)$. The clustering coefficient of the
graph is then given by:
$$C(G) = \frac{1}{N_2}\sum_{\{i \in V| d_i \geq 2\}} c(i),$$
where $N_2$ is the number of nodes with $d_i \geq 2.$

A more global measure of the presence of triangles is called the
\textit{transitivity} of graph $G$ and is defined as:
$$T(G) = \frac{3 \delta(G)}{\tau(G)}.$$ Although they are often similar, $T(G)$ and $C(G)$ can vary by orders of magnitude
\cite{soffer}.  They differ most when the triangles are heterogeneously
distributed in the graph.

\par
These traditional measures of clustering are degree-dependent 
and thus can be biased by the degree sequence of the network.
The maximum number of possible triangles for a given node $i$ 
is just its number of triples ($\tau(i)$). 
For a node which is connected to only low
degree neighbors, however, the maximum number of possible triangles
may be much smaller than $\tau(i)$.  To account for this, a new measure for
clustering was introduced in \cite{soffer} that calculates triadic closure as a 
function of degree and neighbor degree. Specifically, the Soffer-Vasquez clustering coefficient ($\tilde{C}$) and
transitivity ($\tilde{T}$) are given by:
$$\tilde{C}=\frac{\sum_{i|\omega_i > 0}\delta(i)/\omega(i)}{N_{\omega}}$$
$$\tilde{T} = \frac{\sum_{i}\delta(i)}{\sum_{i}\omega(i)},$$
where $\omega(i)$ measures the number of \textit{possible} triangles
for node $i$, and $N_{\omega}$ is the number of nodes in $G$ for
which $\omega(i) > 0$. We note that $\tilde{C}$ and $\tilde{T}$ are undefinited if  $\omega(G)= \sum_i \omega(i)=0$.  $\omega(i)$ is computed by counting the
maximum number of edges that can be drawn among the
$d_i$ neighbors of a node $i$, given the degree sequence of $i$'s neighbors; this value is often smaller than
$\nchoosek{d_i}{2}$ \cite{soffer}. For example, consider a star network of five nodes, where four nodes have degree 1 and one node has degree 4.
Although the total number of triples is $\tau(G)=6$, the number of possible triangles is $\omega(G) = 0$ because the degree one nodes preclude their formation.

\SubSection{Generating Random Graphs} Generating random graphs
uniformly from the set of simply connected graphs with a prescribed
degree sequence is a well-studied problem with algorithmic solutions
\cite{milo}. One of the simplest and most popular of these
generative algorithms was originally suggested by Molloy and Reed
\cite{molloy_reed}. Their model, however, sometimes produces graphs that
are not simple or connected. This can be remedied by subsequently
removing multiple edges and self loops from the constructed graph
and keeping only the largest connected component. Although this
approach works, the Markov Chain Monte Carlo (MCMC) method for
generating simple connected graphs with specified realizable
degree sequences \cite{realizable, erdos} presented
in \cite{gkantsidis,milo} is less prone to problems. It proceeds as follows:
\begin{enumerate}
\item Create a graph with the desired degree sequence using the deterministic Havel-Hakimi algorithm. \cite{havel, hakimi}.
\item Connect any disconnected components of the graph using the
Taylor algorithm \cite{taylor}.
\item Randomly rewire the graph while keeping it
simple and connected \cite{milo}.
\end{enumerate}
The Havel-Hakimi algorithm is iterative and tracks the residual degree of each
node, which is the difference between its current degree and desired degree.
In each iteration, it picks an arbitrary node $x$ and adds edges from $x$ to $d_x$ other nodes with the highest
residual degrees, where $d_x$ is the degree of $x$. The residual degrees of all the nodes are then updated.
The Taylor algorithm merges disconnected components of a graph by randomly selecting edges $(i,j)$ and $(k,l)$
from different components of the graph and rewiring them to $(i,k)$ and $(j,l)$, as long as the rewiring does not
create new disconnected components.

\vspace{0in}

\SubSection{Markov Chain Model} Our method of generating clustered
graphs can be described by a Markov chain. We let $D$ be a
realizable degree sequence and define $G_D$ to be the set of all
simple, connected graphs with degree sequence $D$.  If
$G_1, G_2, ..., G_{|G_D|}$ are the graphs of $G_D$, then we let
$X_1, X_2, ..., X_{|G_D|}$ be the states of the Markov chain, $P$,
where $X_i$ represents the state in which our graph $G = G_i$. The
states $X_i$ and $X_{i+1}$ are connected in the Markov Chain if $G_{i}$ can be changed
to $G_{i+1}$ with the rewiring of one pair of edges. The state space of
the Markov chain $P$ is connected because there exists a path from
$X_i$ to $X_j$ (for any pair $i,j$) by one or more rewiring moves
that leave the degree sequence unchanged \cite{taylor}.

\par
Our clustered graph generation algorithm involves first obtaining a graph,
$G$ of $G_D$ by the method outlined in Section 2.2, and then
transitioning from the state corresponding to $G$ ($X_G$) to other
states of $P$ until a halting condition is reached.  A transition
from one state of the Markov chain to another only happens when the
algorithm makes an edge rewiring that both increases the number of
triangles in the graph and leaves the graph connected. Since a
rewiring does not alter the degree sequence of the graph, the
rewired graph is still in $G_D$. The transition probabilities of the
Markov chain for a pair of connected states, $X_i$ to $X_j$, are:
\[ P_{ij} = \left\{
\begin{array}{ll}
         1 & \mbox{if $(\Delta_j-\Delta_i) > 0$ and $G_j$ is connected}\\
         0 & \mbox{otherwise}\\
\end{array} \right. \]
where $\Delta_i$ is the number of triangles in $G_i$. The algorithm
continues searching for a feasible rewiring (one that increases the
number of triangles and does not disconnect the graph) until one is
found. If a feasible move is not found, a transition is not made and
the process remains in the current state.

\par
The Markov chain above is finite and aperiodic, but not irreducible
as the process can never transition to a state in which the graph
has fewer triangles. It does, however, have an absorbing state,
$X_*$, in which the transitivity of $G_*$ is greater than or equal
to the desired transitivity or is the maximum possible transitivity
given the particular degree sequence and connectivity constraints.

\vspace{-0.25in}
\begin{figure*}[h]
\begin{center}
\includegraphics[width=14cm]{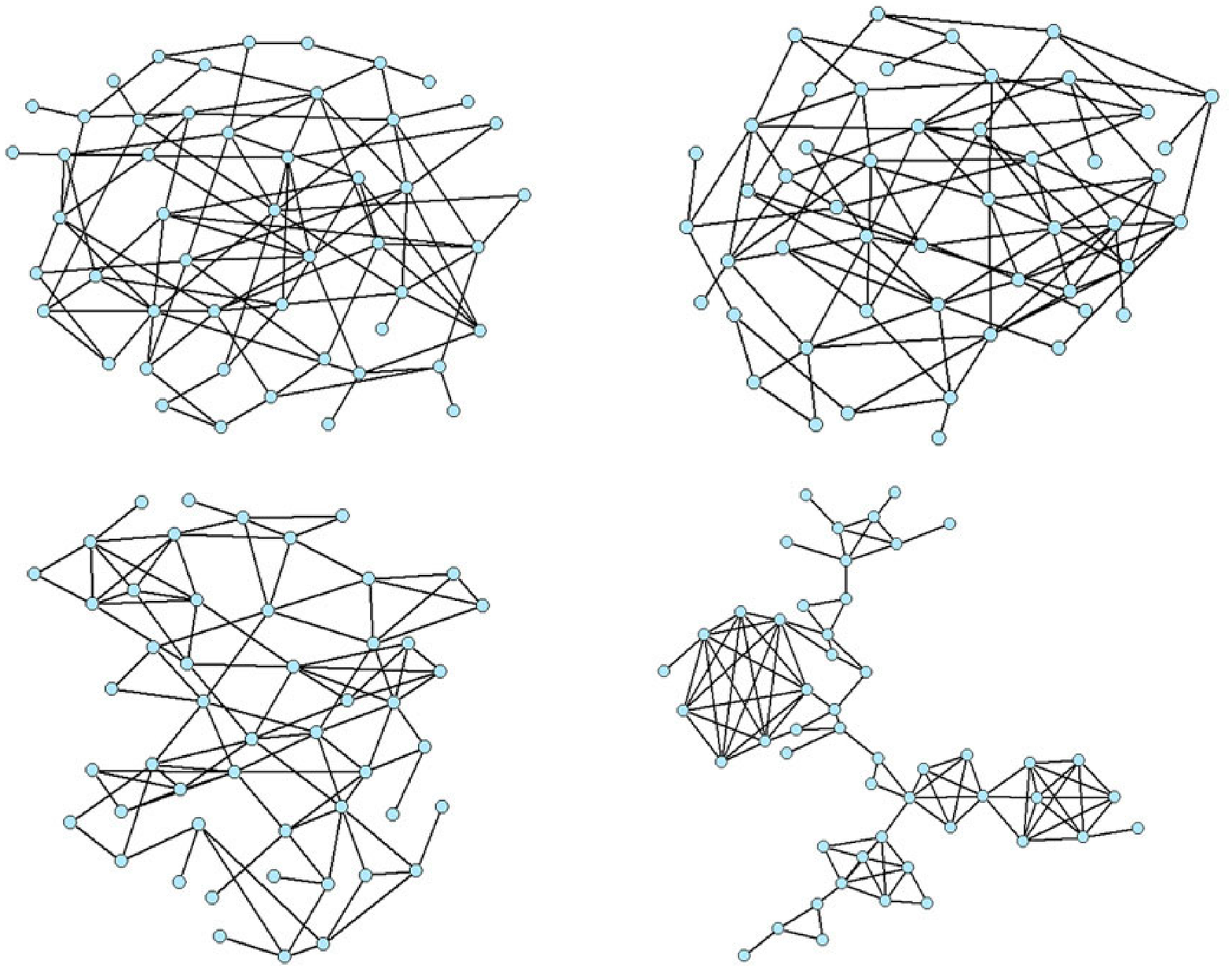} \caption{The evolution of the graph from
(a) $\tilde{T} \approx 0$ ,(b) $\tilde{T} = 0.1$,(c) $\tilde{T} =
0.5$ and (d) $\tilde{T} = 0.8$ \cite{pajek}}
\end{center}
\end{figure*}

\SubSection{Algorithm} Our Markov Chain simulation algorithm for
generating clustered random graphs is described below and
illustrated in Figure 1d.

\begin{algorithm}[H]
  \SetLine
  \KwIn{A random graph, $G$, with a realizable
degree sequence $\{d_i\}$, generated using the method outlined in
Section 2.2 or another suitable method, a desired
clustering value, $target$, and a tolerance value \textit{TOL}.} \KwInit{Measure the
clustering of $G$, $clust(G)$.}
  \While{$|clust(G) - target| \geq$ \textit{TOL}} 
  {
    \begin{enumerate}
    \vspace{0.15in}

    \item uniformly select a random node, $x$, from the\\
    set of all nodes of $G$ such that $d_x > 1$.

    \item uniformly select two random neighbors, $y_1$\\
          and $y_2$, of $x$ such that $d_{y_1}>1$ and \\
          $d_{y_2}>1$ and $y_1 \neq y_2$.

    \item uniformly select a random neighbor, $z_1$\\
          of $y_1$ and a random neighbor, $z_2$ of \\
          $y_2$ such that $z_1 \neq x$, $z_2 \neq x$, \\
          $z_1 \neq z_2$.

    \item $G_{cand} := G$ where $G_{cand}$ is the candidate\\
          graph to which the transition may be made.

    \item \If{$(y_1, y_2)$ and $(z_1,z_2)$ do not exist}{
        \Indp{Rewire two edges of $G_{cand}$: delete $(y_1,z_1)$ and $(y_2,z_2)$, add $(y_1, y_2)$ and $(z_1,
        z_2)$.}
    }

    \item Update the value of $clust(G_{cand})$ by measuring\\
          $\delta(i)$ (and $\omega(i)$ if relevant) for the nodes involved \\
          in the rewiring and their neighbors.

    \item \If{$clust(G_{cand}) > clust(G)$ and $G_{cand}$\\
           is connected}{
        \Indp{$G := G_{cand}$}
    }

    \end{enumerate}
  }
  \KwOut{A random graph, $G$ with degree sequence $\{d_i\}$
and $clust(G) = target \pm $TOL.}
\end{algorithm}

\par
The algorithm terminates when the desired clustering (within a
given tolerance) or the maximum clustering possible is reached. In
the latter case, the desired clustering is not achieved given the
degree and connectivity constraints. 
Theoretically, the algorithm may never reach the target, but if it does, the answer is guaranteed to be correct (this is also sometimes known as a Las Vegas type
algorithm). For practical implementation purposes, a threshold can be placed on the number of iterations run by the algorithm in the case that the desired clustering cannot be reached.

\subsubsection{Choice of Clustering Measure}
The algorithm is defined independent of the
choice of clustering measure. The term $clust(G)$ in the algorithm above
can be replaced by any clustering measure described in Section 2.1, or, more simply, 
the number of triangles in the graph.
\par
The choice of clustering measure does, however, affect the output of the algorithm.
The clustering coefficient is a local measure; and thus $C$ and $\tilde{C}$ yield networks that are only locally optimized for the desired level of clustering. Also, as connectivity is required by our algorithm, the algorithm does not generate graphs
which must be disconnected into multiple components to attain high levels of clustering. (An example of this is given in Appendix Figure 8).
The algorithm may also have difficulty attaining target clustering values when using the standard clustering measures ($C$ or $T$) because of joint degree constraints (the degrees of adjacent nodes) on the possible numbers of triangles, as with the example presented in Section 2.1.  The Soffer-Vasquez clustering measures, which explicitly consider joint degree constraints, provide a way around this difficulty \cite{soffer}.  Although the rewiring in our algorithm changes the joint degree distribution (and thus the degree correlations) of the graph, $\omega(G)$ is not altered significantly
during network generation (as shown in Appendix Figure 9).  Thus, when using $\tilde{C}$ or $\tilde{T}$, clustering is increased primarily by
the addition of triangles (that is, increasing $\delta(G)$) rather than decreasing $\omega(G)$).

\begin{table*}[t]
\begin{center}
{\small
\begin{tabular}{|c|c|c|c|c|c|c|c|c|}
\hline
\\[0.25pt]
Generated Network Type&$N$&$<d>$&$<d^2>$&$T$&$\tilde{T}$&$Diam$&$r$&$Q$\\
\hline
Vancouver Urban Contacts&2627&13.9&265&0.09[0] &0.14 [0]&6 [0]&0.15 [-0.4]&0.28 [-0.15]\\
WWW Subgraph&4271&4.2&119&0.03 [0.02]&0.1 [0.02]&15 [5]&0.07[0.37]&0.45[-0.15]\\
Yeast Protein Interactions&4713&6.3&152& 0.07[0.01]&0.18 [0]&12.5 [3.5]&0.11 [0.07]&0.39 [-0.1]\\
Astro-Phys Collaborations&5973&4.1&35&0.26[-0.05]&0.62[0]&17 [-3]&0.25 [-0.07]&0.70[-0.1]\\
US Air Traffic&165&38.0&2765&0.58[0]&0.97 [0]&3 [0]&-0.55 [0]&0.11 [-0.01]\\
\hline
\end{tabular}
} 
\end{center}
\caption{Comparisons between empirical networks and random graphs. For each empirical network, we generated 25 random graphs constrained to have the observed degree sequences and Soffer-Vasquez transitivity values.  The
table reports average values of several network statistics for the random graphs:
network size ($N$), mean degree ($\langle d \rangle$), mean
squared degree ($\langle d^2 \rangle$), Soffer-Vasquez clustering
coefficient ($\tilde C$), Soffer-Vasquez transitivity ($\tilde T$),
maximum shortest path length between any two nodes ($diam$), degree
correlation coefficient ($r$), and modularity ($Q$). The value given
in brackets is the deviation of the
ensemble mean from the corresponding statistic for the empirical network. (A
positive relative deviation indicates that the ensemble mean was
greater than the empirical statistic and vice versa.) Deviations are
not listed for $N$, $\langle d \rangle$ and $\langle d^2 \rangle$ as 
network size and degree sequence are constrained by our algorithm to match the empirical networks perfectly.}
\end{table*}

\SubSection{Analysis} As shown in Figure 2, there are six types of
triangles that can be added or removed for every pair of edges that
are rewired. As illustrated in Figure 1d, these additions and removals can occur in combination.

Type A: The addition of the edge between vertices $y_1$ and $y_2$
guarantees the addition of one triangle in every rewiring event.

Type B: The addition of the edge $(y_1, y_2)$ could create new
triangles with shared neighbors of $y_1$ and $y_2$.

Type C: The addition of the edge $(z_1, z_2)$ could add a triangle
if there existed edges between $x$ and $z_1$ and $x$ and $z_2$.

Type D: The addition of the edge between vertices $z_1$ and $z_2$
could create new triangles with shared neighbors of $z_1$ and $z_2$.

Type E: The removal of edges $(y_1, z_1)$ and $(y_2, z_2)$ removes
one triangle each if the edges $(x, z_1)$ or $(x, z_2)$ exist.

Type F: The removal of the edges between vertices $y_1$ and $z_1$,
and $y_2$ and $z_2$ could lead to the removal of existing triangles
with shared neighbors of $y_1$ and $z_1$ or $y_2$ and $z_2$.

We note that although the type A addition is a special case of type
B, the type C addition is a special case of type D, 
and the type E removals are a special case of type F, 
we distinguish them because they have different probabilities of occurrence.  
Our look-ahead strategy only allows rewiring moves 
when the total number of Type E and F losses is fewer than the
total number of Type A, B, C, and D gains.

\vspace{-0.15in}
\begin{figure*}[h]
\begin{center}
\includegraphics[width=13cm]{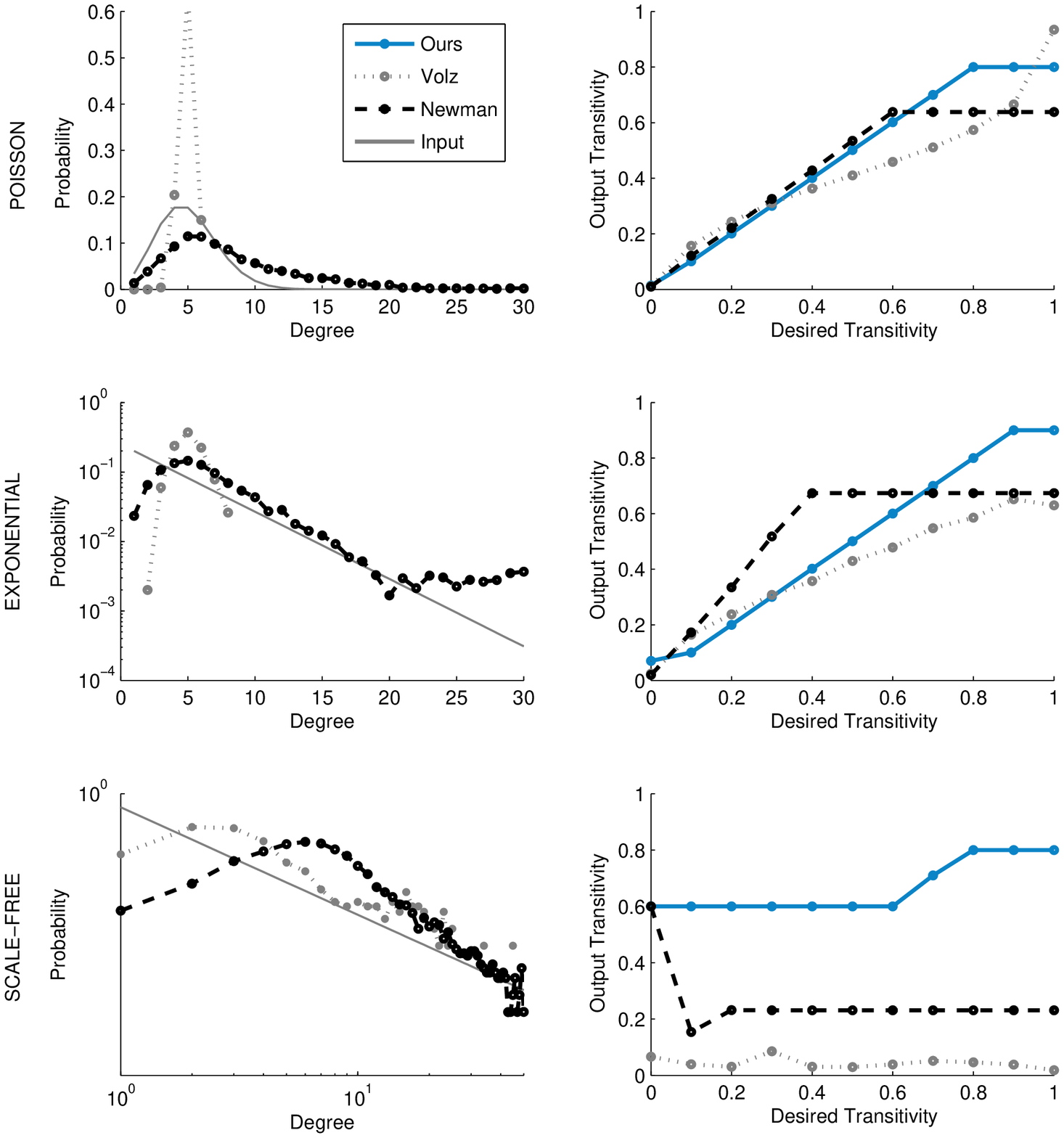}
\caption{Discrepancies between input and output degree
distributions (left panels) and transitivity values (right
panels) for an ensemble of 15 Poisson (top panels), exponential
(middle panels) and scale-free graphs (bottom panels) as generated
by our algorithm and the algorithms presented in \cite{volz_clust} and
 \cite{mejn_clust}. Each graph has $N = 500$ and mean degree, $\langle d \rangle =
5$.  The input degree distribution is shown as a gray solid line (left graphs); and output degree distributions are not shown for our algorithm as they always perfectly match the input. The input and
output transitivity values are measured as $\tilde{T}$ for our
algorithm, and as $T$ for the Volz and Newman algorithm.}
\end{center}
\end{figure*}

\SubSection {Computational Complexity} Like many MCMC methods, the
algorithm we propose can be computationally expensive.  The
method outlined in Section 2.2 requires $O(M)$ steps to generate a
connected graph, and up to $O(M)$ steps to randomize the graph, where $M$ is the number of edges in the graph. At
each step of randomization, we test that the graph remains connected
(an $O(M)$ operation), resulting in an overall $O(M^2)$ network generation process. 
A naive computation of the transitivity/clustering coefficient requires checking every node for the existence of edges between every pair
of neighbors of the node.  This step requires $O(Nd_{max}^2)$ operations,
where $N$ is the number of nodes and $d_{max}$ is the maximum degree of any node in the graph.  

The most expensive step of our algorithm is the introduction of triangles via rewiring. A single rewiring step
requires $O(M)$ operations for switching edges, checking for
connectivity and updating the triangle count.  Although we cannot calculate analytically the number of rewiring steps required to reach
the desired transitivity, we have found it empirically to be
$O(M)$. Thus, the average complexity of the algorithm presented here
is $O(M^2)$. This complexity has been computed for the most naive
versions of our algorithms; and more efficient
implementations may improve the complexity greatly. For example, we might improve efficiency by
performing connectivity tests once
every $x$ rewirings (for some number $x$) rather than during every rewiring, as
proposed in \cite{gkantsidis}.

\vspace{-0.05in}
\begin{figure*}[h]
\begin{center}
\includegraphics[width=15cm]{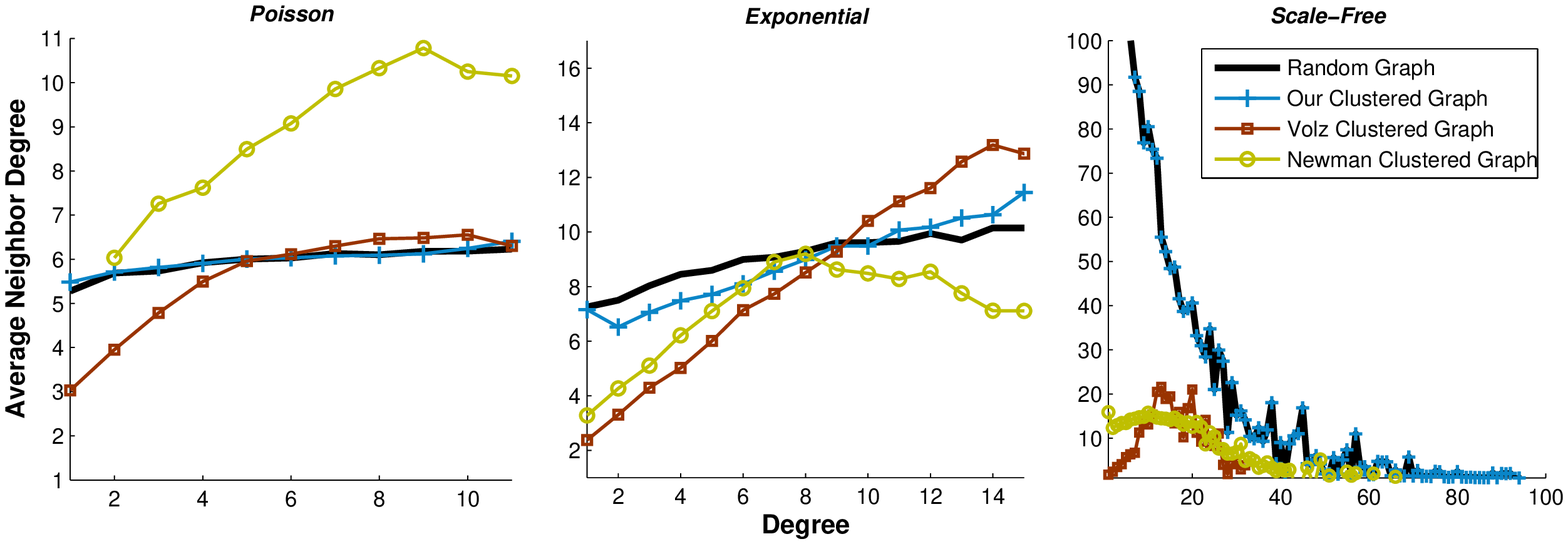}
\caption{Degree correlations in random graphs with
specified degree distributions (Poisson, exponential and scale-free
with mean degree = 5) compared to clustered random graphs with the same degree
distributions and $\tilde{T} = 0.5$ generated by our algorithm and the Volz \cite{volz_clust} and Newman
\cite{mejn_clust} algorithms. The graphs present averages over
15 graphs generated by each algorithm. Our algorithm introduces
fewer degree correlations than the alternatives.}
\end{center}
\end{figure*}

\Section{Results}

\SubSection{Numerical Simulations}

To check the feasibility and reliability of the algorithm, we generated networks for several
degree distributions and a range of clustering values. Specifically, we used Poisson $\left(p_d = e^{-\lambda}\lambda^d/d!\right)$,
exponential $\left(p_d = (1-e^{\kappa})e^{-\kappa (d-1)}\right)$ and
scale-free $\left(p_d = d^{-\gamma}/\zeta(\gamma)\right)$ degree
distributions, and Soffer-Vasquez transitivity ($\tilde{T}$) ranging from 0
to 1 in increments of 0.1. Figure 3 illustrates a
graph (N=50) with a Poisson distributed degree sequence evolving
towards higher transitivity.

\par
We evaluated the performance of the algorithm in comparison to those proposed in \cite{volz_clust} (as
a representative of the growing networks class of clustered graphs
generators) and in \cite{mejn_clust} (as a representative of the
class of bipartite models). Specifically we measured the discrepancies between input and output degree distributions (Figure 4, left graphs) and transitivity values (Figure 4, right graphs).
Our algorithm preserves the input degree sequence perfectly, while there are considerable mismatches between the input and output degree
distributions in the Volz and Newman models.
For Poisson and exponentially distributed graphs, our algorithm
closely approaches the target transitivity. These degree distributions
cannot, however, reach the highest transitivity values (Figures 4b and 4d) without disconnecting the graph.  
Unlike our algorithm, the Volz and Newman models
do not require connectivity, which may explain the superior performance of the Volz algorithm on the Poisson network at the maximum transitivity value (Figure 4b).
The Volz algorithm also performs well at low values of $T$ for both the Poisson and exponential networks (Figures 4b and 4d);
while the Newman algorithm only performs well on the Poisson networks.

\par
Our algorithm performs quite poorly on scale-free random graphs (Figure 4f), which have much higher
clustering a priori than expected for Poisson random graphs
\cite{serrano, park}. Our algorithm is not designed to decrease clustering, and therefore can only reach the desired level if the initial random graph has lower clustering than desired.  The triangles in a connected scale-free random graph are also close to the minimum required to keep the graph connected, and thus, modifying our algorithm to decrease (as well as increase) the
triangles in a graph would likely not improve its performance on the scale-free graphs. 

\SubSection{Structural Properties of Our Generated Networks}

There are several other topological
properties (besides degree sequence and transitivity) that can strongly influence network function and dynamics: degree correlations (the
dependence of a node's degree on its neighbor's degrees), community structure (groups of nodes that are highly intra-connected and only loosely inter-connected), and average path length (typical distances between pairs of nodes in the network). We have specifically developed this model to increase clustering with minimal structural byproducts. Thus, we confirm that we have reached this goal by measuring the above properties in the networks generated by our algorithm.

We evaluated the extent to which the algorithm introduces degree correlations by comparing random (unclustered) graphs to clustered random graphs generated by our algorithm and the Volz \cite{volz_clust} and Newman \cite{mejn_clust}
 algorithms (Figure 5). While our algorithm essentially preserves the correlation structure of the random graph, the other algorithms produce highly correlated graphs.

\par
Several authors have discussed the relationship between clustering
and community structure \cite{park, radicchi, ravasz, serrano}. As
Figure 3 shows, the addition of triangles leads to modular structure. This
behavior is not surprising: as the number of edges in the graph is
constrained, sets of connected nodes with high $\omega(i)$ values
(often high-degree nodes) must be brought together to create
additional clustering.

\par
Short average path lengths are a characteristic feature of random graphs
\cite{mejn_rand}. To quantify the impact of our algorithm on path lengths, we calculated the average path length for each node to all other $(N-1)$ nodes, and then compared the distributions of these values for several random and random clustered graphs (Figure 6).  While our algorithm preserves short average path
lengths, the mean of the path length distribution tends to be slightly larger
for the clustered graphs than for the corresponding random graphs.  The intuition behind this increase in average path length may lie in the increased community structure: as graphs become more clustered and separate into subgroups, nodes in different groups require more links to reach each other.

\vspace{-0.2in}
\begin{figure*}[h]
\begin{center}
\includegraphics[width=13cm]{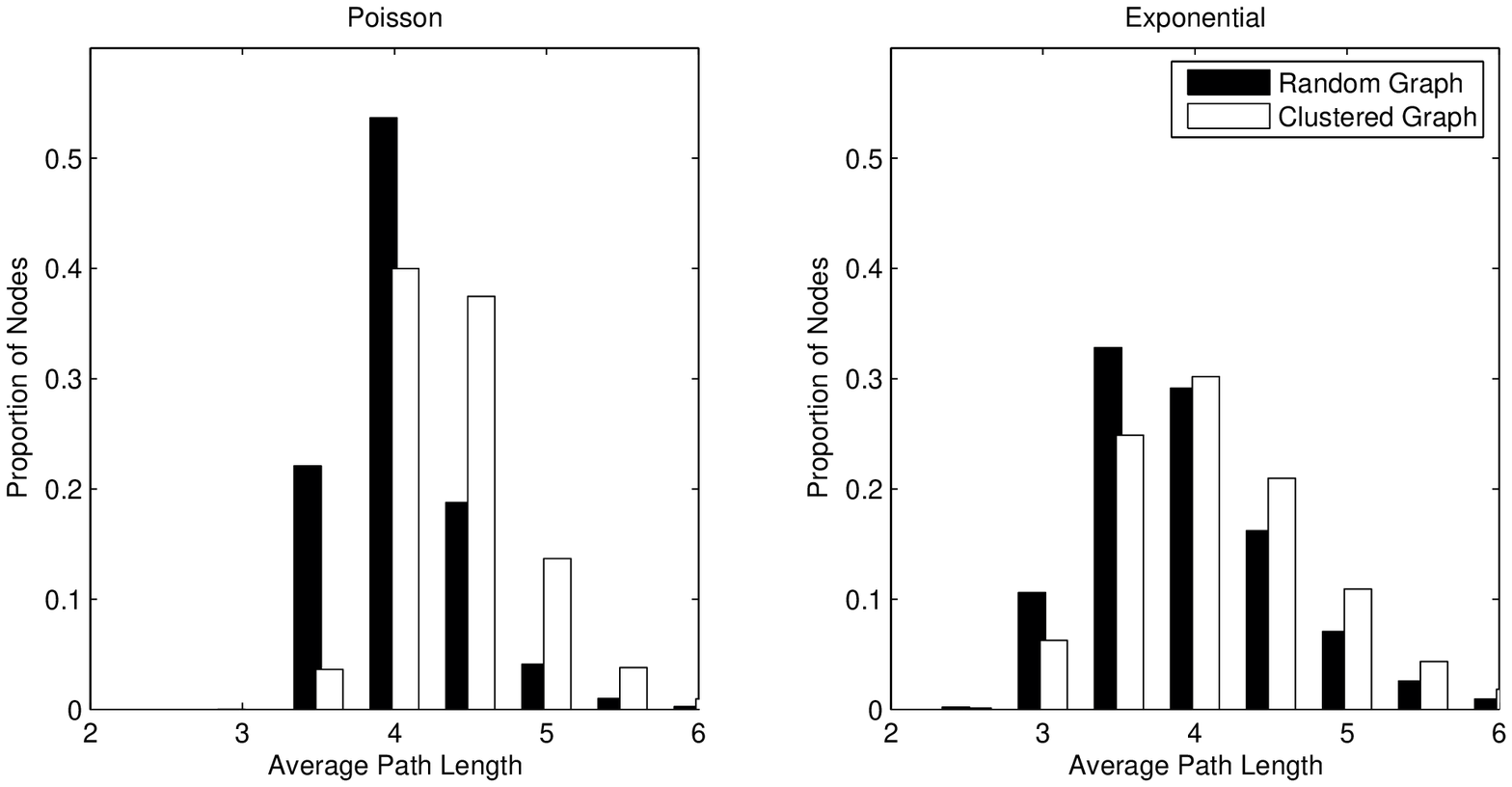}
\caption{Average path lengths in
random graphs with specified degree distributions (Poisson and
exponential with mean degree = 5) compared to clustered random graphs with the same
degree distributions and $\tilde{T} = 0.5$ generated by our
algorithm. The histograms are based on 15 networks of each type. The clustered graphs have slightly higher means than their
random counterparts: 4.05 for the Poisson random graphs verus 4.39 for the clustered graphs; and 3.95 for the exponential random
graphs versus 4.14 for the custered graphs.}
\end{center}
\end{figure*}

\SubSection{Comparison to Empirical Networks} 
Our algorithm can also be applied to detecting non-random structure in empirical 
networks. We can generate ensembles of clustered random networks with empirically estimated degree distributions and clustering values to ascertain whether empirical networks have significant non-random structure in other respects.
We demonstrate this application using representatives from five different classes of real networks: (1) a social network, made up of contacts
between individuals in the city of Vancouver \cite{meyers_sars}, (2)
a protein interaction network for yeast \cite{yeast}, (3) a
technological network, made up of a subset of the links of the World
Wide Web \cite{www_graph}, (4) a transportation network, made up of
US metropolitan areas connected by air travel \cite{airline_data},
and (5) a collaboration network, made up of scientists connected by
coauthorship on scientific preprints on the Astrophysics E-Print
Archive between 1995 and 1999 \cite{astro}, with a collaboration
strength of 0.5 or greater \cite{mejn_collab}. The basic statistics
of these networks, including clustering values, are listed in Table
1.

\par
We used the following method to quantify deviations from
randomness in these networks. First, we used our algorithm to generate 25 clustered random networks constrained to match the empirical degree distribution and clustering values. Second, we selected a set
of network topological measures (other than degree distribution and
clustering), and compared these quantities for the empirical graph to the corresponding average quantities across the ensemble of generated graphs.

\par
Specifically, we generate 25 random clustered networks for each empirical network, constrained to match the empirical degree sequence and Soffer-Vasquez transitivity.
In addition to the degree and clustering metrics, we also calculated diameter (longest
shortest path length between any pair of nodes in the graph) \cite{harary}, degree correlation coefficient \cite{mejn_ass} and modularity
(degree of community structure) \cite{mejn_cs} (Table 2). Other than diameter, each of these range from 0 to 1. The
standard deviations for all statistics are negligible and thus 
not reported. For every statistic, we also give the deviation between the empirical value and the average across the generated ensemble of random clustered networks (specifically, deviation = ensemble mean - observed
value). Small 
deviations suggest that the empirical network structure
boils down to the degree distribution and clustering, and thus we turn our attention to possible mechanisms underlying these properties.
In contrast, large deviations suggest that there are other fundamental properties to consider in addition to or, perhaps, instead of clustering.

\par
The random counterparts of the US air traffic
network, for example, have structural properties almost identical to
the real network, suggesting that the structure of the US air
traffic network comes almost exclusively from its degree patterns. (In fact, even the high clustering is explained exclusively by the degree patterns.)
We note that the US air traffic network
is the most engineered of the networks we consider, and thus may have fewer emergent properties.
The remaining empirical networks
differ considerably from their random counterparts, suggesting
that there are important mechanistic features not captured in our random model. For example, the two social networks (the Vancouver urban contact
network and the Astro-Phys collaboration network) have higher degree assortativity than our random networks. This may point to rules of social behavior
beyond that dictated by number of ``friends'' and the tendency that ``my friend's friend is also my friend.''  All the natural networks also have significantly higher community structure than the corresponding random networks, inspite of having a wide range of transitivity values. This shows that clustering and community structure are not necessarily postively correlated.

\vspace{-0.08in}
\Section{Conclusion} In this work, we have introduced a Markov chain
simulation algorithm to generate clustered random graphs with a
specified degree sequence and level of clustering. Our algorithm
perfectly preserves the degree sequence of a random graph and generally maintains
other fundamental properties of random graphs like short path length and low degree
correlations. An ensemble of the graphs generated by this algorithm
can thus be useful for systematically studying the impact of triangles on network function and dynamics and understanding identifying the essential structural features of empirical networks. Since this method is based on a dynamic process, it can be used to generate both static networks with a specified amount of clustering and dynamic networks with evolving levels of clustering. Furthermore, since the
process is a ``memoryless" one, additional clustering can be added to any network without having to grow a new one from scratch.

\section*{Acknowledgements}
The authors acknowledge valuable feedback from Mark Newman, Erik
Volz, Alberto Segre and Ted Herman; and support for L.A.M. from the McDonnell Foundation.

\bibliographystyle{plain}  
\bibliography{clust}

\begin{thebibliography}{10}
\bibitem{realizable}
A realizable degree sequence is one which satisfies the handshake theorem (the
  requirement that the sum of the degrees be even) and the erdos-gallai
  criterion (which requires that for each subset of the $k$ highest degree
  nodes, the degrees of these nodes can be ``absorbed'' within the subset and
  the remaining degrees.
\end{thebibliography}

\section*{Appendix}
We evaluated the effectiveness of an algorithm which accepts
all rewirings regardless of their effect on the number of triangles. Recall that our main algorithm only makes rewirings that
increase the number of triangles.  In Figure 7, we show that the permissive algorithm is
not effective in achieving the desired levels of clustering. Additionally, other structural properties of the network, e.g. degree correlations,
are significantly altered from the orginal graph in this case (as shown in the right panel of Figure 7.)

Figure 8 illustrates a network in which disconnection is required to achieve maximal clustering. 

Figure 9 shows that our algorithm does not change
the number of possible triangles ($\omega(G)$) in the graph drastically.
\vspace{-0.25in}
\begin{figure*}[h]
\begin{center}
\includegraphics[width=13cm]{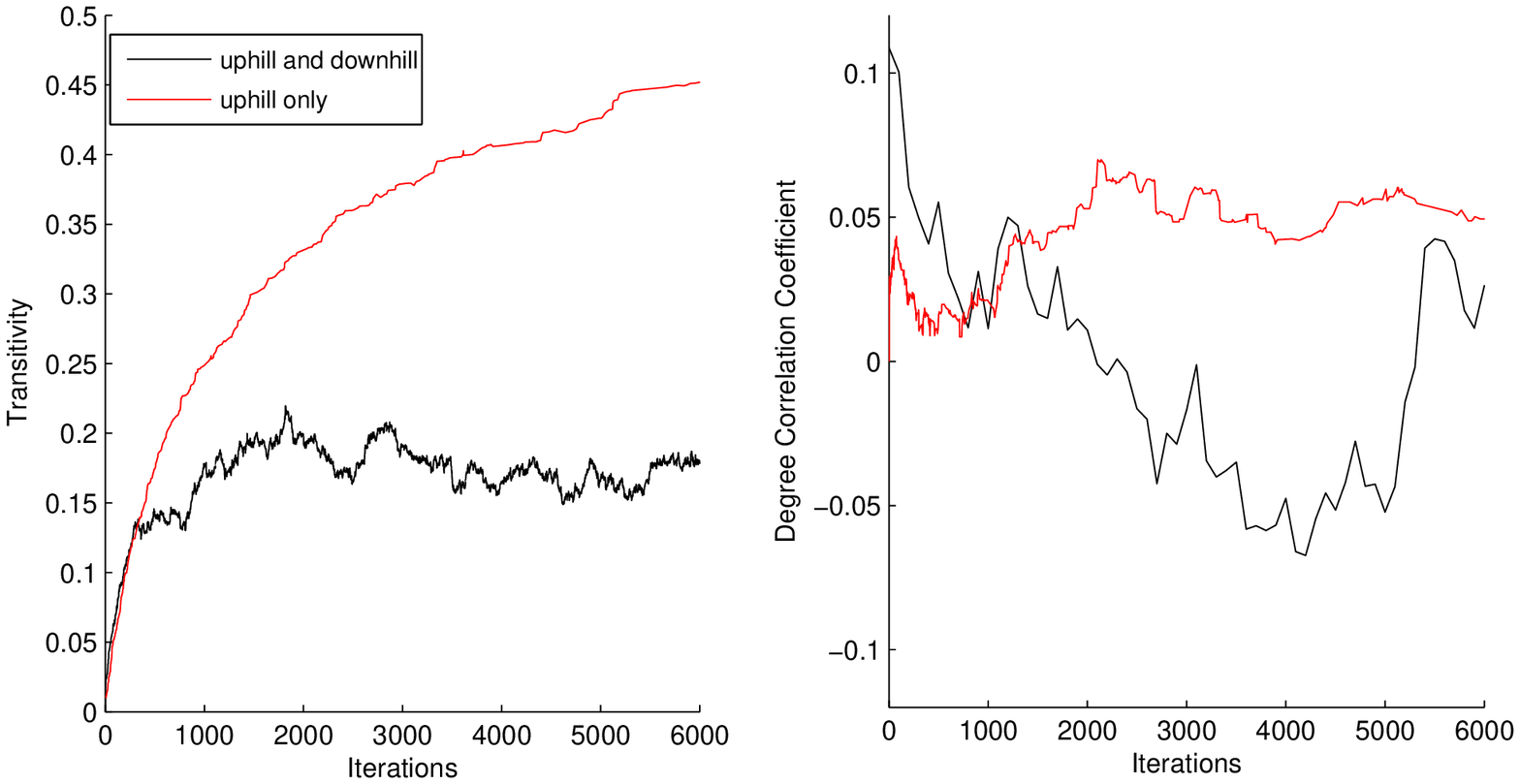}
\end{center}
\caption{The effect of allowing both uphill (rewirings that increase
the total number of triangles) and downhill (rewirings that
decrease the total number of triangles) moves. These results are shown for
a Poisson distributed graph of 500 nodes. In the left panel, we see that allowing all rewirings is not effective
in reaching the desired transitivity ($T = 0.45$).  We also find that the structure of the graph is altered significantly in the process
of making all rewirings. The degree correlation coefficient, for example, varies significantly with each rewiring (as shown in the 
right panel.}
\end{figure*}

\begin{figure*}[h]
\begin{center}
\includegraphics[width=18cm]{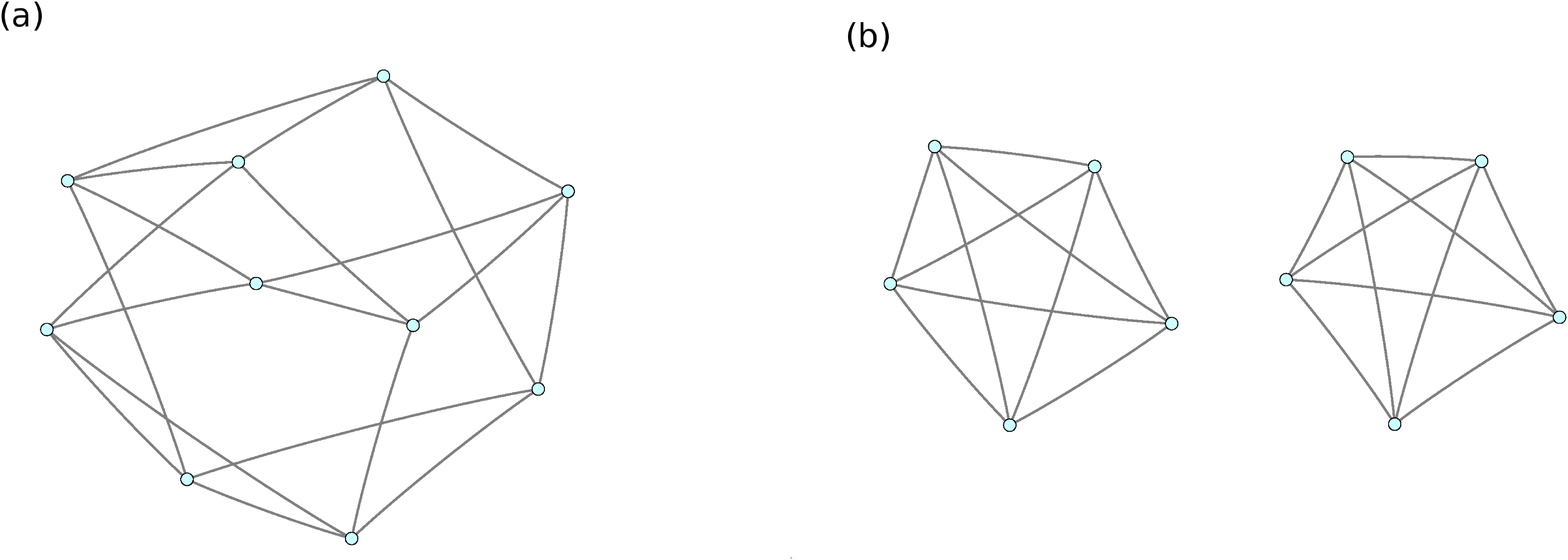}
\caption{(a) A random graph with 10 nodes, each of degree 4. (b) The graph in (a) must be disconnected 
to be maximally clustered ($C = T = \tilde{C} = \tilde{T} = 1$). }
\end{center}
\end{figure*}

\begin{figure*}[h]
\begin{center}
\includegraphics[width=13cm]{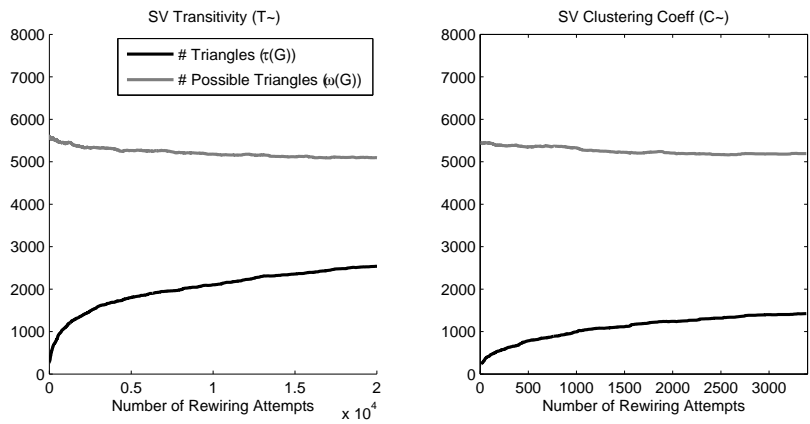}
\caption{The numbers of triangles $\delta(G)$ and possible triangles $\omega(G)$ in the graph as
the algorithm progresses.  $\omega(G)$
does not vary substantially during graph generation. These results are for a Poisson
distributed graph of 500 nodes, to which triangles are added until reaching (a) Soffer-Vasquez transitivity = 0.5
and (b) Soffer-Vasquez clustering coefficient = 0.5.}
\end{center}
\end{figure*}

\end{document}